\documentclass[twocolumn,english,superscriptaddress, amsmath,amssymb,notitlepage,aps,twocolumn,prl]{revtex4-2}
\usepackage[T1]{fontenc}
\usepackage[latin9]{inputenc}
\usepackage{amsbsy}
\usepackage{amstext}
\usepackage{graphicx}

\makeatletter

\providecommand{\tabularnewline}{\\}

\usepackage[colorlinks,linkcolor=blue, urlcolor=blue, anchorcolor=blue, citecolor=blue]{hyperref}
\usepackage{dcolumn}
\usepackage{bm}

\makeatother

\usepackage{babel}
\begin{document}
\title{Measurement Uncertainty in Infrared Spectroscopy with Entangled Photon
Pairs}
\author{Xue Zhang}
\address{Graduate School of China Academy of Engineering Physics, Beijing 100193,
China}
\author{Zhucheng Zhang}
\email{zczhang@gscaep.ac.cn}

\address{Graduate School of China Academy of Engineering Physics, Beijing 100193,
China}
\address{School of physics, Hangzhou Normal University, Hangzhou 310036, China}
\author{Hui Dong}
\email{hdong@gscaep.ac.cn}

\address{Graduate School of China Academy of Engineering Physics, Beijing 100193,
China}
\date{\today}
\begin{abstract}
Spectroscopy with entanglement has shown great potential to break
limitations of traditional spectroscopic measurements, yet the role
of entanglement in spectroscopic multi-parameter joint measurement,
particularly in the infrared optical range, remains elusive. Here,
we find an uncertain relation that constrains the precision of infrared
spectroscopic multi-parameter measurements using entangled photon
pairs. Under such a relation, we demonstrate a trade-off between the
measurement precisions of the refractive index and absorption coefficient
of the medium in the infrared range, and also illustrate how to balance
their respective estimation errors. Our work shall provide guidance
towards the future experimental designs and applications in entanglement-assisted
spectroscopy. 
\end{abstract}
\maketitle
\textit{Introduction}.---Entanglement, as a non-classical resource,
plays an important role in advancing various domains of physics \citep{Horodecki2009}.
This unique quantum phenomenon facilitates correlations between particles
that defy classical explanation, enabling a range of applications
previously deemed impossible. Entanglement\textquoteright s utility
spans multiple areas, including quantum computation \citep{jozsa2003role,penrose1998quantum,brassard1998quantum,deutsch1985quantum},
where it enhances computational power beyond classical limits; quantum
cryptography \citep{Bennett2014,Gisin1999,Duer1999}, providing unparalleled
security measures; high-precision metrology \citep{PhysRevLett.96.010401,giovannetti2011advances,pezze2018quantum,DemkowiczDobrzanski2014,riedel2010atom},
improving measurement accuracy; and quantum communication \citep{Cleve1997,Buhrman2001,Brukner2004,ursin2007entanglement,gisin2007quantum,yuan2010entangled},
enabling faster and more secure data transmission. Moreover, the exploration
of entanglement\textquoteright s potential in other fields is ongoing
\citep{casacio2021quantum,PhysRevLett.130.133602,PhysRevLett.125.180502,li2023single,brida2010experimental,kalashnikov2016infrared,okamoto2020loss,zhang2024quantum},
continuously revealing new possibilities and driving forward the frontier
of quantum technology.

One particularly promising application of entanglement is in the realm
of spectroscopic measurements \citep{RevModPhys.54.697,Jackson1995,Gremlich2000,Stuart2004,solarz2017laser},
which are crucial for providing information on the compositions and
structural characteristics of a wide range of samples, from complex
materials to delicate biological tissues. In this regard, entanglement
has been utilized to assist infrared measurement and to construct
new infrared spectroscopic measurement schemes \citep{kalashnikov2016infrared,paterova2018measurement,paterova2017nonlinear}.
Traditional infrared measurement typically requires infrared laser
as well as the corresponding detectors, which often need cryogenic
cooling to reduce thermal noise to improve sensitivity and accuracy
\citep{Rogalski2000,Rogalski2003,Karim2013,Keyes2013}. By exploiting
the unique correlations between entangled photons, particularly those
generated through spontaneous parametric down-conversion (SPDC), it
has been demonstrated that the absorption coefficient and refractive
index of the medium in the infrared range can be determined from the
measurements of visible photons, so that the limitations of traditional
infrared detection methods can be bypassed \citep{kalashnikov2016infrared}.
Despite some experimental attempts, the role of entanglement in improving
the precision of spectroscopic multi-parameter measurements remains
unclear, especially in the context of infrared spectroscopy. 

In this Letter, we establish an inherent uncertainty relation that
imposes precision limits on infrared spectroscopic multi-parameter
measurements using entangled photon pairs. Unlike single-parameter
measurements, the optimal measurements for multiple parameters typically
faces incompatibility due to the constraints imposed by the Heisenberg's
uncertainty principle \citep{heisenberg1983physical,busch2007heisenberg}.
This measurement incompatibility fundamentally limits the achievable
precision in infrared spectroscopic multi-parameter measurements \citep{PhysRevLett.126.120503,xia2023toward}.
With such uncertainty relation, we demonstrate a trade-off between
the measurement precisions of refractive index and absorption coefficient,
which inherently limits the precision achievable in their joint measurement.

\begin{figure}
\centering{}\includegraphics{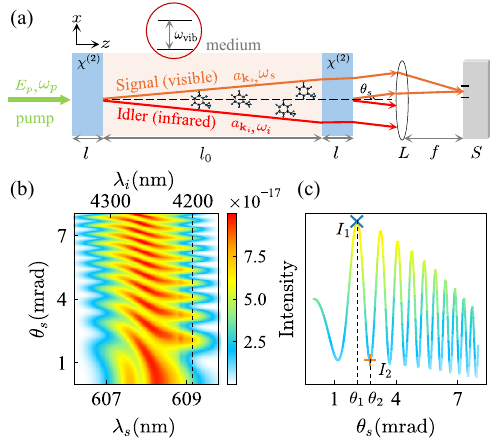}\caption{Setup for entanglement-assisted infrared spectroscopic measurement.
(a) The measurement system includes two nonlinear crystals with a
quadratic susceptibility $\chi^{(2)}$ and the same length $l$, separated
by a chamber of length $l_{0}$. The two crystals are pumped by a
continuous-wave laser with amplitude $E_{p}$ and angular frequency
$\omega_{p}$ (green line) to induce the SPDC process, generating
entangled photon pairs with visible-signal photons at frequency $\omega_{s}$
(orange line) and infrared-idler photons at frequency $\omega_{i}$
(red line). During the first SPDC process, the entangled signal-idler
photons propagate towards the chamber and interact with the medium
that vibrates at infrared frequency $\omega_{\mathrm{vib}}$ in the
chamber. After the two SPDC processes, the entangled photon pairs
from the two crystals are imaged by a lens ($L$) onto a visible-light
spectrometer ($S$) positioned at the lens\textquoteright s focal
length ($f$). (b) Simulated two-dimensional wavelength-angle intensity
distribution of the output signal light, where the wavelength $\lambda_{i}$
of the idler photons is inferred from the energy conservation, i.e.,
$\lambda_{i}^{-1}=\lambda_{p}^{-1}-\lambda_{s}^{-1}$. (c) A cross-section
of the two-dimensional wavelength-angle intensity at the fixed wavelength
of $609.16$ nm for signal photons. The two intensities $I_{1}$ and
$I_{2}$ are the peak and valley in the output intensity spectrum
of the signal light at emission angle $\theta_{1}=2.09$ mrad and
$\theta_{2}=2.73$ mrad, respectively. Other parameters used in the
numerical simulation are given in Table \ref{tab:table1}.\protect\label{fig:system}}
\end{figure}

\textit{Entanglement-assisted infrared spectroscopic measurement}.---We
consider the setup of entanglement-assisted infrared spectroscopic
measurement \citep{kalashnikov2016infrared}, as depicted in Fig.~\ref{fig:system}(a).
In this setup, a strong pump field with amplitude $E_{p}$ and angular
frequency $\omega_{p}$ from a continuous-wave laser is applied to
a nonlinear crystal of length $l$ to induce the SPDC process, in
which the quadratic susceptibility $\chi^{(2)}$ describes the strength
of the interaction between the pump field and the crystal. This crystal
is configured to generate entangled photon pairs with signal photons
in the visible optical range at frequency $\omega_{s}$, and idler
photons in the infrared optical range at frequency $\omega_{i}$.
Under a weak SPDC process, the quantum state of these signal-idler
photons is \citep{SM}
\begin{eqnarray}
\left|\psi\right\rangle  & \approx & (1+\sum_{\mathbf{k}_{s}\mathbf{k}_{i}}F_{\mathbf{k}_{s}\mathbf{k}_{i}}a_{\mathbf{k}_{s}}^{\dagger}a_{\mathbf{k}_{i}}^{\dagger})\left|00\right\rangle ,
\end{eqnarray}
where $a_{\mathbf{k}_{s}}^{\dag}(a_{\mathbf{k}_{i}}^{\dag})$ is the
creation operator of signal (idler) light with wave vector $\mathbf{k}_{s}(\mathbf{k}_{i})$,
and $F_{\mathbf{k}_{s}\mathbf{k}_{i}}$ is the probability amplitude
for creating a pair of signal and idler photons with specific wave
vectors $\mathbf{k}_{s}$ and $\mathbf{k}_{i}$. The entangled signal-idler
photons then propagate towards a chamber of length $l_{0}$, in which
the medium (e.g., $\mathrm{CO_{2}}$ gas with infrared vibration frequency
$\omega_{\mathrm{vib}}$) is transparent to visible light and absorbs
light in the infrared region, namely \citep{RevModPhys.84.621},
\begin{eqnarray}
a_{\mathbf{k}_{s}} & \rightarrow & e^{-\mathrm{i}\phi_{s}}a_{\mathbf{k}_{s}},
\end{eqnarray}
and
\begin{eqnarray}
a_{\mathbf{k}_{i}} & \rightarrow & \tau e^{-\mathrm{i}\phi_{i}}a_{\mathbf{k}_{i}}+\mathrm{i}\sqrt{1-\tau^{2}}v,
\end{eqnarray}
with $v$ as the annihilation operator of the vacuum environment.
Here, $\phi_{j}=\mathbf{k}_{j}\cdot\mathbf{d}_{j}+\mathbf{k}_{j}^{m}\cdot\mathbf{d}_{j}^{m}\,(j=s,i)$
are the phases introduced by the propagation of the signal and idler
lights in the crystal and medium. $\mathbf{k}_{j}^{m}$ are the wave
vectors in the medium, with magnitudes $|\mathbf{k}_{j}^{m}|=|\mathbf{k}_{j}|n_{j}^{m}/n_{j}$,
where $n_{j}$ and $n_{j}^{m}$ are the refractive indices of the
crystal and medium, respectively. $\mathbf{d}_{j}$ and $\mathbf{d}_{j}^{m}$
are the distance vectors in the crystal and the medium. And $\tau=\exp(-\alpha l_{0})$
is the amplitude transmissivity of the medium for the idler photons,
with $\alpha$ as its amplitude absorption coefficient.

After propagating through the medium, the pump field interacts with
an identical crystal and induces a second SPDC process, generating
additional entangled photon pairs. These entangled photons from the
two crystals are imaged by a lens ($L$) onto the input slit of a
visible-light spectrometer ($S$), which is positioned at the lens's
focal length ($f$). The slit of the spectrometer allows signal photons
with an emission angle $\theta_{s}$ to enter and separates them based
on different wavelengths $\lambda_{s}$, forming a two-dimensional
wavelength-angle intensity distribution. This two-dimensional intensity
distribution is determined by the average number of photons of the
output signal light \citep{klyshko1993ramsey,SM}, i.e., 
\begin{eqnarray}
I(\lambda_{s},\theta_{s}) & \equiv & \left\langle \psi_{\mathrm{out}}\right|a_{\mathbf{\mathbf{k}_{s}}}^{\dagger}a_{\mathbf{\mathbf{k}_{s}}}\left|\psi_{\mathrm{out}}\right\rangle \nonumber \\
 & \propto & \left[\mathrm{sinc}\left(\frac{\delta}{2}\right)\right]^{2}[1+\tau\cos(\delta+\delta^{m})],\label{eq:intensity}
\end{eqnarray}
where
\begin{eqnarray}
\left|\psi_{\mathrm{out}}\right\rangle  & \approx & \left|\boldsymbol{0}\right\rangle -\mathrm{i}\sum_{\mathbf{k}_{s}\mathbf{k}_{i}}F_{\mathbf{k}_{s}\mathbf{k}_{i}}\sqrt{1-\tau^{2}}e^{\mathrm{i}\phi_{s}}a_{\mathbf{k}_{s}}^{\dagger}v^{\dagger}\left|\boldsymbol{0}\right\rangle \nonumber \\
 &  & +\sum_{\mathbf{k}_{s}\mathbf{k}_{i}}F_{\mathbf{k}_{s}\mathbf{k}_{i}}\tau e^{\mathrm{i}(\phi_{i}+\phi_{s})}a_{\mathbf{k}_{s}}^{\dagger}a_{\mathbf{k}_{i}}^{\dagger}\left|\boldsymbol{0}\right\rangle \nonumber \\
 &  & +\sum_{\mathbf{k}_{s}\mathbf{k}_{i}}F_{\mathbf{k}_{s}\mathbf{k}_{i}}e^{\mathrm{i}\phi_{p}}a_{\mathbf{k}_{s}}^{\dagger}a_{\mathbf{k}_{i}}^{\dagger}\left|\boldsymbol{0}\right\rangle \label{eq:output=000020state}
\end{eqnarray}
is the output state of system, with $\left|\boldsymbol{0}\right\rangle \equiv|000\rangle$
as the vacuum state for the signal, idler and environment modes. The
phase term, $\phi_{p}=k_{p}l+k_{p}^{m}l_{0}$, is introduced by the
propagation of the pump field, in which $k_{p}$ and $k_{p}^{m}$
are the wave vectors of the pump field within the crystal and the
medium, respectively. Here, $\delta=l\Delta$ and $\delta^{m}=l_{0}\Delta^{m}$
are the phase mismatches that arise due to the wave vector mismatches
$\Delta=k_{p}-k_{sz}-k_{iz}$ and $\Delta^{m}=k_{p}^{m}-k_{sz}^{m}-k_{iz}^{m}$
in the crystal and the medium, with $k_{jz}$ and $k_{jz}^{m}$ as
the wave vectors of the signal-idler photons along the $z$ direction,
respectively. The dependence of intensity $I(\lambda_{s},\theta_{s})$
on emission angle $\theta_{s}$ and wavelength $\lambda_{s}$ is reflected
in the expression for the phase mismatches $\delta$ and $\delta^{m}$,
whose explicit forms are given in the Supplementary Materials \citep{SM}.
These two phase mismatches ultimately lead to the occurrence of the
interference pattern observed in the visible-light spectrometer.

\begin{table}[t]
\caption{\protect\label{tab:table1}System parameters used in the numerical
simulation \citep{kalashnikov2016infrared,paterova2017nonlinear,paterova2018measurement}.}

\begin{ruledtabular}
\begin{tabular}{lr}
Parameters & Values\tabularnewline[0.1cm]
\hline 
\noalign{\vskip0.1cm}
Wavelength of the pump light $\lambda_{p}$ & 532 nm\tabularnewline
Quadratic susceptibility $\chi^{(2)}$ & 20 pm/V\tabularnewline
Crystal length $l$ & 0.5 mm\tabularnewline
Chamber length $l_{0}$ & 25 mm\tabularnewline
Amplitude absorption coefficient $\alpha$ & 0.15 $\mathrm{cm}^{-1}$\tabularnewline
Refractive indexes $n_{p}^{m}/n_{s}^{m}/n_{i}^{m}$ & 1/1/$1-9\times10^{-5}$\tabularnewline
Refractive indexes $n_{p}/n_{s}/n_{i}$ & 2.3232/2.2930/2.1052\tabularnewline
\end{tabular}
\end{ruledtabular}

\end{table}

The output intensity of the signal light is illustrated as functions
of its wavelength $\lambda_{s}$ and emission angle $\theta_{s}$
in Fig.~\ref{fig:system}(b), with system parameters provided in
Table \ref{tab:table1}. It is clear that the output intensity exhibits
distinct bright and dark stripes within the visible wavelength range
of the signal light, as shown by the dashed line in Fig.~\ref{fig:system}(b)
and its corresponding cross-section in Fig.~\ref{fig:system}(c).
By measuring the output intensities of the signal light with different
emission angles and then fitting these measurement data using Eq.
(\ref{eq:intensity}), the refractive index $n_{i}^{m}$ and absorption
coefficient $\alpha$ of the medium for the infrared-idler photons
are inferred simultaneously in the experiment \citep{kalashnikov2016infrared}. 

\textit{Precision of joint measurement of refractive index and absorption
coefficient.---}The output intensity of the visible-signal light
is correlated with the refraction and absorption properties of the
medium for the infrared-idler light, enabling the joint measurement
of these two parameters. By selecting two intensities, $I_{1}$ and
$I_{2}$, corresponding to different emission angles {[}e.g., $\theta_{1}$
at the peak and $\theta_{2}$ at the valley, as illustrated in Fig.~\ref{fig:system}
(c){]} in the output intensity spectrum of the signal light as observables,
the joint measurement precisions of the refraction index $n_{i}^{m}$
and absorption coefficient $\alpha$ are determined by the following
covariance matrix \citep{paris2009quantum,barlow1993statistics,SM},
\begin{equation}
\boldsymbol{\mathcal{C}}=\left(\begin{array}{cc}
\sigma_{n_{i}^{m}}^{2} & \textrm{cov}(n_{i}^{m},\alpha)\\
\textrm{cov}(\alpha,n_{i}^{m}) & \sigma_{\alpha}^{2}
\end{array}\right),
\end{equation}
in which $\sigma_{n_{i}^{m}}^{2}=M^{-1}\sum_{k\in\{1,2\}}(\partial n_{i}^{m}/\partial I_{k})^{2}\sigma_{I_{k}}^{2}$
and $\sigma_{\alpha}^{2}=M^{-1}\sum_{k\in\{1,2\}}(\partial\alpha/\partial I_{k})^{2}\sigma_{I_{k}}^{2}$
are their measurement variances, and $\textrm{cov}(n_{i}^{m},\alpha)=M^{-1}\sum_{k\in\{1,2\}}(\partial n_{i}^{m}/\partial I_{k})(\partial\alpha/\partial I_{k})\sigma_{I_{k}}^{2}=\textrm{cov}(\alpha,n_{i}^{m})$
is covariance, respectively. Here, $\sigma_{I_{k}}^{2}\equiv\left\langle \psi_{\mathrm{out}}\right|(a_{\mathbf{k}_{s}}^{\dagger}a_{\mathbf{k}_{s}})^{2}\left|\psi_{\mathrm{out}}\right\rangle -\left\langle \psi_{\mathrm{out}}\right|a_{\mathbf{\mathbf{k}_{s}}}^{\dagger}a_{\mathbf{\mathbf{k}_{s}}}\left|\psi_{\mathrm{out}}\right\rangle ^{2}$
is the variance of the output signal photons with specific wave vector
$\mathbf{k}_{s}$, and $M$ is the number of repeated measurements
in the experiment. Additionally, the precision of the joint measurement
is bounded by the quantum Cramér-Rao inequality in parameter estimation
theory \citep{cramer1999mathematical,fisher1923xxi,helstrom1969quantum,holevo2011probabilistic,PhysRevLett.72.3439,liu2020quantum},
that is,
\begin{equation}
\boldsymbol{\mathcal{C}}\geq(M\boldsymbol{\mathcal{F}})^{-1},\label{eq:QCRB}
\end{equation}
in which the matrix $\boldsymbol{\mathcal{F}}$ with elements
\begin{eqnarray}
\mathcal{\boldsymbol{\mathcal{F}}}_{\mu,\nu} & \equiv & 4\mathrm{Re}(\langle\partial_{\mu}\psi_{\mathrm{out}}|\partial_{\nu}\psi_{\mathrm{out}}\rangle\nonumber \\
 &  & -\langle\partial_{\mu}\psi_{\mathrm{out}}|\psi_{\mathrm{out}}\rangle\langle\psi_{\mathrm{out}}|\partial_{\nu}\psi_{\mathrm{out}}\rangle)
\end{eqnarray}
is the quantum Fisher information matrix, representing the maximum
information about the refractive index and absorption coefficient
that can be extracted from all measurement methods. $\mathrm{Re}(\bullet)$
denotes the real part, $\langle\partial_{\mu}\psi_{\mathrm{out}}|$
and $\langle\partial_{\nu}\psi_{\mathrm{out}}|$ are abbreviations
for $\partial\langle\psi_{\mathrm{out}}|/\partial\mu$ and $\partial\langle\psi_{\mathrm{out}}|/\partial\nu$,
with $\mu,\nu\in\{n_{i}^{m},\alpha\}$, respectively.

\begin{figure}[t]
\centering{}\includegraphics{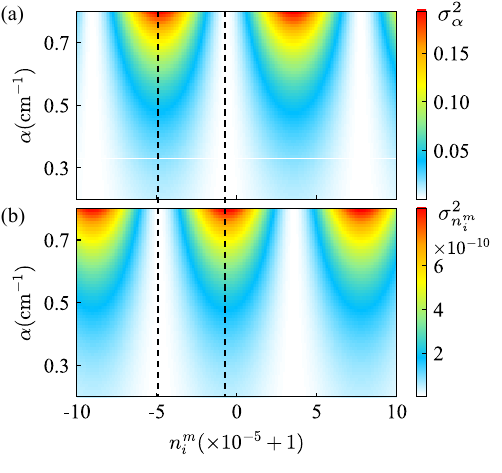}\caption{Dependencies of the variances $\sigma_{\alpha}^{2}$ (a) and $\sigma_{n_{i}^{m}}^{2}$
(b) on refractive index $n_{i}^{m}$ and absorption coefficient $\alpha$.
The color in the plots represents different variance values as indicated
by the color bar. Here, considering a $10\:\mathrm{s}$ pump duration
and a $10^{3}\:\mathrm{s}^{-1}$ biphoton flux \citep{kalashnikov2016infrared,PhysRevLett.124.163603},
the number of repeated measurements $M$ is set as $10^{21}$ to obtain
cumulative photon numbers of the same magnitude on the spectrometer.
And other parameters are the same as in Fig.~\ref{fig:system} and
Table \ref{tab:table1}.\protect\label{fig:Delta}}
\end{figure}

Figures \ref{fig:Delta} (a) and (b) illustrate the dependence of
the measurement variances $\sigma_{n_{i}^{m}}^{2}$ and $\sigma_{\alpha}^{2}$
on refractive index $n_{i}^{m}$ and absorption coefficient $\alpha$,
respectively. We demonstrate that when $\alpha$ is fixed, the two
variances exhibit oscillatory behavior as $n_{i}^{m}$ varies; whereas,
for a constant $n_{i}^{m}$, they attain a smaller value in the range
of weak absorption $\alpha$. It is important to mention that the
positions of the refractive index for the two minimum variances $\sigma_{n_{i}^{m}}^{2}$
and $\sigma_{\alpha}^{2}$ are different. Specifically, the refractive
index that minimizes $\sigma_{\alpha}^{2}$ approximately corresponds
to the refractive index that maximizes $\sigma_{n_{i}^{m}}^{2}$,
and vice versa, as shown by the dashed lines in Figs.~\ref{fig:Delta}
(a) and (b). This implies that when jointly measuring the refractive
index and absorption coefficient, there is a trade-off relation between
their measurement precisions, which prevents them from reaching their
minimum values simultaneously.

\begin{figure}[b]
\centering
\centering{}\includegraphics{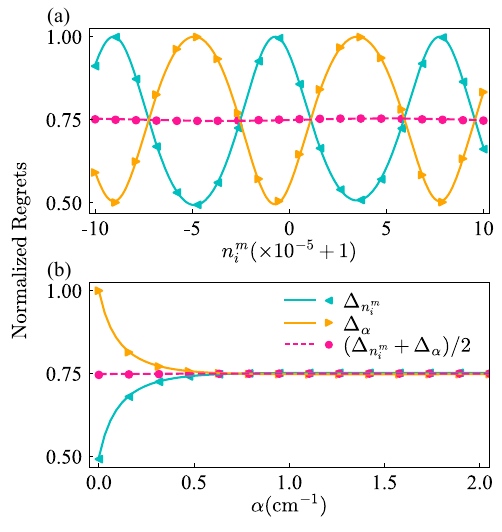}\caption{Normalized Fisher information regrets $\Delta_{n_{i}^{m}}$ and $\Delta_{\alpha}$
versus refractive index $n_{i}^{m}$ and absorption coefficient $\alpha$,
respectively, where the solid and dashed curves represent these regrets
based on their definitions, while the corresponding triangle and dot
symbols are drawn from their approximate expressions {[}i.e. Eqs.
(\ref{eq:nm})-(\ref{eq:relation}){]}. Here, (a) $\alpha=0.2\,\mathrm{cm^{-1}}$,
(b) $n_{i}^{m}=1-7\times10^{-5}$, and other parameters are the same
as in Fig.~\ref{fig:system} and Table \ref{tab:table1}. \protect\label{fig:bound}}
\end{figure}

\textit{Trade-off relation}.---To reveal the trade-off relation between
the measurement precisions of refractive index and absorption coefficient,
we here define the regret \citep{PhysRevLett.126.120503} of Fisher
information based on Eq.~(\ref{eq:QCRB}), namely,
\begin{equation}
\boldsymbol{\mathcal{R}}\equiv\boldsymbol{\mathcal{F}}-\boldsymbol{\mathcal{C}}^{-1},
\end{equation}
in which the repeat measurement number $M$ is set to $1$ without
loss of generality. This Fisher information regret characterizes the
gap between the quantum Fisher information and the information extracted
from the joint measurement regarding these two parameters to be estimated.
By introducing the normalized regret of Fisher information, i.e.,
$\Delta_{\mu}=\boldsymbol{\mathcal{R}}_{\mu\mu}/\boldsymbol{\mathcal{F}}_{\mu\mu}$,
we demonstrate that for the case of close emission angles ($\theta_{1}\approx\theta_{2}$),
the normalized regrets of Fisher information for the refractive index
and absorption coefficient in their joint measurement are derived
as \citep{SM}
\begin{eqnarray}
\Delta_{n_{i}^{m}} & \approx & 1-\sum_{k\in\{1,2\}}\frac{\sin^{2}(\delta_{k}+\delta_{k}^{m})}{4[1+\tau\cos(\delta_{k}+\delta_{k}^{m})]},\label{eq:nm}
\end{eqnarray}
and
\begin{eqnarray}
\Delta_{\alpha} & \approx & 1-\sum_{k\in\{1,2\}}\frac{(1-\tau^{2})\cos^{2}(\delta_{k}+\delta_{k}^{m})}{4[1+\tau\cos(\delta_{k}+\delta_{k}^{m})]},\label{eq:alpha}
\end{eqnarray}
respectively. Here $\delta_{k}\equiv\delta(\theta_{k})$ and $\delta_{k}^{m}\equiv\delta^{m}(\theta_{k})$
are the phase mismatches corresponding to $\theta_{k}$. We find that
the two normalized regrets of Fisher information satisfy the following
trade-off relation,
\begin{equation}
\Delta_{n_{i}^{m}}+\Delta_{\alpha}\approx\frac{3}{2}+\sum_{k\in\{1,2\}}\frac{\tau\cos(\delta_{k}+\delta_{k}^{m})}{4}.\label{eq:relation}
\end{equation}
Meanwhile, considering the two intensities $I_{1}$ and $I_{2}$ that
correspond to the peak and valley in the output intensity spectrum
of the signal light, Eq.~(\ref{eq:intensity}) indicates that the
term $\cos(\delta_{1}+\delta_{1}^{m})\approx1$ and the term $\cos(\delta_{2}+\delta_{2}^{m})\approx-1$.
This leads to a simplified trade-off relation, namely, $\Delta_{n_{i}^{m}}+\Delta_{\alpha}\approx3/2$.
Additionally, when the two intensities correspond to different peaks,
the trade-off relation becomes as $\Delta_{n_{i}^{m}}+\Delta_{\alpha}\approx3/2+\tau/2$.
And when the two intensities correspond to different valleys, the
relation is $\Delta_{n_{i}^{m}}+\Delta_{\alpha}\approx3/2-\tau/2$.
This clearly demonstrates that for strong absorption media with transmissivity
$\tau\approx0$, regardless of whether the peak or valley in the output
signal intensity is chosen as the observable, the lower bound of the
trade-off is approximately $3/2$. And for weak absorption media with
transmissivity $\tau\approx1$, selecting the output intensities corresponding
to different valleys as observables can potentially yield higher joint
measurement precisions and a lower bound of the trade-off with relation
$\Delta_{n_{i}^{m}}+\Delta_{\alpha}\approx1$. However, it is impossible
to simultaneously maximize the estimation precisions of these two
parameters (i.e., achieving $\Delta_{n_{i}^{m}}=\Delta_{\alpha}=0$)
under physically permissible scenarios.

The trade-off relation in Eq.~(\ref{eq:relation}) indicates that
when jointly measuring the refractive index and absorption coefficient
of a medium for idler light, if more information about the refractive
index is extracted by measuring the output intensity of the signal
light, then correspondingly, less information about the absorption
coefficient will be extracted, and vice versa. To illustrate this
relation, the normalized Fisher information regrets $\Delta_{n_{i}^{m}}$
and $\Delta_{\alpha}$ are plotted as functions of $n_{i}^{m}$ and
$\alpha$ with the intensities at peak and valley as observables,
respectively, as shown in Figs.~\ref{fig:bound}(a) and (b). The
solid and dashed curves depict the exact values of these normalized
Fisher information regrets based on their definitions, whereas the
corresponding triangle and dot symbols are their approximate values
obtained from Eqs.~(\ref{eq:nm})-(\ref{eq:relation}). From the
curves, we can see that as $n_{i}^{m}$ and $\alpha$ vary, the two
normalized Fisher information regrets exhibit a mutually constraining
behavior, while their sum remains approximately around $3/2$, illustrated
with the two dashed lines in figures. It is clear that under this
trade-off relation, the estimation precision of one parameter is constrained
by the measurement inaccuracy of the other, showing that the estimation
precisions of the refractive index and absorption coefficient cannot
be maximized simultaneously.

\textit{Conclusion}.---By introducing multi-parameter estimation
theory into spectroscopic measurements, our study reveals the inherent
uncertainty in multi-parameter measurements within infrared spectroscopy.
This uncertainty imposes precision limits on the joint measurement
of multiple parameters, not only in the field of linear spectroscopic
measurement, but also potentially in high-order spectroscopic measurements
\citep{mukamel1995principles}. We have demonstrated that when using
entangled photon pairs to jointly measure the refractive index and
absorption coefficient in the infrared optical range, a trade-off
relation exists between the precisions of these two estimated parameters.
Such trade-off relation holds significant implications for the future
design of high-precision spectroscopic techniques and quantum sensors,
guiding advancements in these fields by highlighting the fundamental
limits of spectroscopic multi-parameter measurements.
\begin{acknowledgments}
We thank Dr.~Lei Shao for helpful discussions. This work is supported
by the Innovation Program for Quantum Science and Technology (Grant
No. 2023ZD0300700), the National Natural Science Foundation of China
(Grant Nos.~U2230203, U2330401, 12088101, 12347123, 12405012), and
the Hangzhou Joint Fund of the Zhejiang Provincial Natural Science
Foundation of China under Grant No. LHZSD24A050001.
\end{acknowledgments}

\bibliographystyle{apsrev4-2}
\bibliography{Refs}

\end{document}